\documentclass[sigconf,nonacm]{acmart}
\setcopyright{none}

\usepackage{amsmath}
\usepackage{amsfonts}
\usepackage{amsthm}
\usepackage{bm}
\usepackage{algorithm}
\usepackage{algpseudocode}

\usepackage{xcolor}
\usepackage{float}
\usepackage{placeins}
\usepackage{graphicx}
\usepackage{booktabs}
\newtheorem{definition}{Definition}
\newtheorem{lemma}{Lemma}

\usepackage{fancyvrb}
\usepackage{xcolor}
\usepackage{tikz}
\usepackage{enumitem}

\newcommand{\Space}[1]{}
\newcommand{\CodeIn}[1]{\begin{small}\texttt{#1}\end{small}}

\fvset{commandchars=\\\{\}}

\newbox\verbbox

\definecolor{Red}{HTML}{D1D1D1}
\definecolor{Green}{rgb}{0.0, 0.56, 0.0}
\definecolor{Gray}{gray}{0.85}
\definecolor{LightBlue}{cmyk}{0.51, 0.1, 0, 0}
\definecolor{Maroon}{cmyk}{0.32, 1.0, 1.0, .46}

\newcommand{\Blue}[1]{\textcolor{blue}{\textbf{#1}}}

\newcommand{\Green}[1]{\textcolor{Green}{\textbf{#1}}}

\newcommand{\ATilde}{\raisebox{0.5ex}{\texttildelow}}

\begin{document}

\title{AlloyASG: Alloy Predicate Code Representation as a Compact Structurally Balanced Graph}


\author{Guanxuan Wu}
\email{gxw6804@mavs.uta.edu}
\orcid{0009-0006-8102-7329}
\author{Allison Sullivan}
\email{allison.sullivan@uta.edu}
\orcid{0000-0001-7400-2218}
\affiliation{%
  \institution{University of Texas at Arlington}
  \streetaddress{P.O. Box 19015}
  \city{Arlington}
  \state{Texas}
  \country{USA}
  \postcode{76019}
}




%



\begin{abstract}
Writing declarative models has numerous benefits, ranging from automated reasoning and correction of design-level properties before systems are built to automated testing and debugging of their implementations after they are built. Unfortunately, the model itself needs to be correct to gain these benefits. Alloy is a commonly used modeling language that has several existing efforts to repair faulty models automatically. Currently, these efforts are search-based methods that use an Abstract Syntax Tree (AST) representation of the model and do not scale. One issue is that ASTs themselves suffer from exponential growth in their data size due to the limitation that ASTs will often have identical nodes separately listed in the tree.
To address this issue, we introduce a novel code representation schema, Complex Structurally Balanced Abstract Semantic Graph (CSBASG), which represents code as a complex-weighted directed graph that lists a semantic element as a node in the graph and ensures its structural balance for almost finitely enumerable code segments. 
We evaluate the efficiency of our CSBASG representation for Alloy models in terms of it's compactness compared to ASTs, and we explore if a CSBASG can ease the process of comparing two Alloy predicates. Moreover, with this representation in place, we identify several future applications of CSBASG, including Alloy code generation and automated repair. 
\end{abstract}
\maketitle

%

\section{Introduction}\label{sec:intro}

Alloy \cite{alloy} is a declarative, relational logic-based modeling language for describing the structural and dynamic properties of a system. Since Alloy is used to verify software system designs~\cite{Zave15,BagheriETAL2018,WickersonETAL2017,ChongETAL2018},
and to perform various forms of analyses over the corresponding implementation, including deep static checking~\cite{JacksonVaziri00Bugs,TACOGaleottiETALTSE2013}, systematic testing~\cite{MarinovKhurshid01TestEra}, data structure repair~\cite{ZaeemKhurshidECOOP2010}, automated debugging~\cite{GopinathETALTACAS2011} and to synthesize security attacks~\cite{websecurity,Margrave,CheckMateMicro2019}, the consequences could be catastrophic if an Alloy model is faulty yet successfully compiles. Several iterative search methods based on ASTs have already been implemented for automated repair and partial generation of Alloy models~\cite{amra, bounded, atrrepair, tsra6}. However,  any iterative search methods for mutants or AST elemental changes that fix the Alloy code segments require expansive storage and have intractable search spaces due to the exponential growth of the code segment length, limiting their computational efficiencies. 

In recent years, there have been several bodies of work that focus on automatically repairing code using machine learning techniques \cite{mlfix1, mlfix3, mlfix4, mlfixs1, mlfixs2, mlfixs3, mlfixs4}. Such works provide valuable inspiration to us since we have a dataset of real-world fixing pairs consisting of faulty predicates written by students and their correct counterparts \cite{alloy4fun}, effectively leading to a set of pairs of inputs and ground truths. Nonetheless, almost all the sequence-based algorithms treat the code segments as natural language sequences with a black-box function trained for faulty code as input and corrected code as output. These approaches ignore the strict logic of the target programming language, severely hindering the explainability of the process and limiting their applicability to modeling languages like Alloy.

To create a machine learning algorithm that is aware of the logic of the programming language we are generating code in, we need a logical structure-aware, vectorized, machine-and-human-readable, and computationally efficient code representation that improves upon the AST adjacency matrix representation that grows exponentially in size. Some of the representation learning methods \cite{code2seq, ast2vec} train functions that output vectors from the AST, yet the nature of their dependency on the control flow paths in the AST of those general-purpose languages renders their methodologies implausible for a specification language such as Alloy, which is declarative and lacks control flow. Other methods \cite{astnn, uastrl, tree2graph} try to focus on the raw AST for a universal representation, yet those algorithms still suffer from rapid growth in the size of the AST. Therefore, this paper aims to create a structure that generates a compact representation of the code segment yet still gives the same complete information as the raw AST. 

Recent work created an Abstract Semantic Graph (ASG) \cite{DuffyASG}, which is close to meeting our data representation needs. Still, the original designation lacks a numerical representation that provides an adjacency or Laplacian matrix that could be used in popular graph machine learning algorithms \cite{gcn1, gcn2, gcn3, graphgan, relationalgcn}. Therefore, in this paper, we introduce a matrix representation of the ASG with the following objectives: 
\begin{enumerate}
    \item One-on-one correspondence between the source code, the AST, and the graph matrix;
    \item Better performance in compactness and the potential to save space compared to the raw AST representation;
    \item Allow a direct comparison of the differences between code segments before and after applying a fix.
\end{enumerate}

Such goals could be impractical for general-purpose, high-level languages such as C or Java, as most languages have a complicated type system and data structures that are not finite or even integer-enumerable. Despite that, for a simple, declarative, almost finitely enumerable language such as Alloy, we could create a Complex-weighted Structurally Balanced Abstract Semantic Graph (CSBASG) that translates the code segments into compact graphs for the subset covering most of the possible combinations. 

One of the advantages of this graph representation is the uniqueness of the semantics. For two semantically different terms in an AST, e.g., a variable or a unary or binary operator, there will only be one corresponding vertex in the CSBASG. This intuitively represents the code elements and their relationships with a reduced space cost. Besides, for a fixing pair of code segments consisting of the faulty code and its correction, the CSBASG could align the modifications by the common node signatures shared by the pair and compare their structural differences. 

Therefore, in this paper, we make the following contributions: 

\newenvironment{Contributions}{}{} 
\newcommand{\Contribution}[1]{\noindent\textbf{#1:}} 

\begin{Contributions}
\Contribution{CSBASG} We introduce the concept of a CSBASG that meets our three objects (one-to-one correspondence, more compact, and supports direct comparisons). We provide an algorithm to convert an AST into a CSBASG and convert a CSBASG back to an AST and corresponding code segments.

\Contribution{Application to Alloy} We tailor the generation of CSBASG to the Alloy modeling language. 

\Contribution{Evaluation} We evaluate the improvement on compactness and density of a CSBASG compared to an AST and evaluate how to compare multiple CSBASGs.

\Contribution{Open Source} We release our source code at \url{https://anonymous.4open.science/r/AlloyASG-Release-CF0F/} (to be replaced with public GitHub in the final version).

\Contribution{Future Directions} We present a series of research directions now enabled due to the representation of Alloy models with a CSBASG.

\end{Contributions}
\FloatBarrier
\section{Background}\label{sec:bg}
This section describes key components of Abstract Semantic Graphs and the Alloy modeling language.

\subsection{Abstract Syntax Tree and Abstract Semantic Graph}
We use the common notation of the abstract syntax tree (AST) for the raw representation of the syntactic structure of the Alloy source code \cite{ast}. We follow the formal definitions from \cite{autoinference}: 
\begin{definition}[Context-Free Grammar]\cite{autoinference}
    A context-free grammar (CFG) $G\equiv (N,\Sigma, R,s)$ is a tuple where $N$ is the set of nonterminals, $\Sigma$ is the set of terminals, $R$ is the set of production rules $a\to b_1b_2...b_N$ where $a\in N$ is a nonterminal and $b_{i}\in N\cup \Sigma$, and $s\in N$ is the starting nonterminal of the grammar. The \textit{language} of $G$ is the set of strings from the starting non-terminal.
\end{definition}
\begin{definition}[AST]\cite{autoinference}
    Given a context-free grammar (CFG) $G=(N,\Sigma, R,s)$, an abstract syntax tree (AST) $T\equiv (G, X, r, \xi, \sigma)$ is a tree with a set of nodes $X$, a root node $r\in X$, a mapping $\xi: X\to \mathcal P(X)$ from each of the nodes to a subset of nodes as its children and each child set $\forall x\in X:\xi(x)$ is either empty or well-ordered, and $\sigma: X\to (N\cup \Sigma)$ maps each node to its corresponding label as a terminal or nonterminal word in the CFG. 
\end{definition}
According to \cite{DuffyASG}, an abstract semantic graph (ASG) resolves the branches defining the properties of a node in the AST to its original definitions. But here, to create a more compact form of an ASG specifically for Alloy, we use a more radical definition that combines every syntactically and semantically equivalent node as below: 
\begin{definition}[ASG]
    Given a CFG $G = (N, \Sigma, R, s)$, an abstract semantic graph (ASG) $\mathcal G\equiv (G, \mathcal V,\mathcal E, r)$ where $\mathcal V\subseteq N\cup \Sigma$ is a set of nonterminal and terminal words as nodes (vertices) of the graph, and $\mathcal E\subseteq \mathcal V\times \mathcal V\times \mathbb C$ is the set of edges between the words with the edge weight between nodes $v_i, v_j\in \mathcal V$ be $w_{ij}\in \mathbb C$, and $r$ is the root of the corresponding AST of the ASG, i.e. the first node being visited. 
\end{definition}

Note that since $\mathcal E$ is an arbitrary set of edges in forms of $(v_i, v_j, w_{ij})$, an AST or empirical construction of ASG could have more than one set of $\{w_{ij}\}$ to represent. If the weight function $w:\mathcal V\times \mathcal V\to \mathbb C$ ensures that $w_{ij}\neq 0$ if and only if there exists an edge from $v_i$ to $v_j$, and $w_{ij}\neq w_{ik}$ for any $j\neq k$ be children of $i$, either at the same order of execution (children under the same node in the corresponding AST) or at the different order (if a nonterminal presents more than once in the AST and each time comes with a set of children). There could also be multiple AST links connecting two ASG nodes, and we will discuss how we form a weight value for this in Section 3. 

\subsection{Complex-weighted Graph and its Structural Balance}
Since \textit{Definition 3} gives a complex-weighted graph and our requirement for a vectorization of such ASG, the construction of the graph requires the edge values specifying the relationships between any of the two nodes. So we borrow the following structural balance concept from \cite{sbg}: 
\begin{definition}[Complex-Weighted Structural Balanced Graph] \cite{sbg}
For a graph $\mathcal G = (\mathcal V, \mathcal E)$, let $a_{ij}=w_{ji}$ be the entries of the adjacency matrix $A$ where $|a_{ij}|>0$ indicates there exists an edge between $v_j$ and $v_i$, then the graph $\mathcal G$ is said to be 
structurally balanced if all the entries of its adjacency matrix
$A = [a_{ij}]^{N\times N}$ satisfies 
$a_{ij} \equiv |a_{ij}|\angle \theta_{ij} = |a_{ij} |\angle(\theta_i - \theta_j)$, where $\theta_1, ..., \theta_N\in (-\pi, \pi)$ is called the \textit{signatures} of the nodes $v_1...v_N$, respectively.
\end{definition}

In the ASG context, intuitively, the signatures of the nodes $\theta_1, ...,\theta_n$ could encode the syntactic and semantic properties of each terminal or nonterminal node and the magnitudes of each edge $|a_{ij}|_{i,j\in \{1...N\}}$ encodes their relative properties, such as if a node is the left (1st) or right (2nd) child of a binary operator. We chose this concept for its mathematical solid properties. Let $d_i=\sum_{j\in 1}^N |a_{ij}|$ which is the in-degree of node $v_i$ and let $D=diag(d_1, ..., d_N)$ be the diagonal matrix giving the in-degrees of each node on their corresponding rows, and define $L=D-A$ as the Laplacian matrix, then we have the lemma below \cite{sbg}:
\begin{lemma}\cite{sbg, sbgcomplete}

\begin{figure*}
\begin{center}
\begin{tabular}[t]{c|c|c}
\footnotesize (a) & \footnotesize (b) & \footnotesize (c) \\

\begin{minipage}[t]{.53\columnwidth}
\scriptsize
\begin{Verbatim}[]
1. \Blue{sig} Person  \{
2.   Tutors : \Blue{set} Person,
3.   Teaches : \Blue{set} Class
4. \}
5. \Blue{sig} Group {}
6. \Blue{sig} Class  \{ Groups : Person -> Group \}
7. \Blue{sig} Teacher \Blue{in} Person  \{\}
8. \Blue{sig} Student \Blue{in} Person  \{\}
\end{Verbatim}
\end{minipage}

&

\begin{minipage}[t]{.72\columnwidth}
\scriptsize
\begin{Verbatim}[]
\Green{/* Assuming a universe of 3 persons, the tutoring }
\Green{ * chain of every person eventually reaches a Teacher. */}
 9. \Blue{pred} inv15oracle \{
10.   \Blue{all} p:Person | \Blue{some} Teacher&(^Tutors).p
11. \}
\end{Verbatim}
\end{minipage}

& 

\begin{minipage}[t]{.5\columnwidth}
\scriptsize
\begin{Verbatim}[]
12. \Blue{pred} inv15 \{
13.   \Blue{all} p : Person \{ \Blue{some} t : Teacher \{
14.     t \Blue{in} p.Tutors 
15.     \Blue{or} t \Blue{in} p.Tutors.Tutors 
16.     \Blue{or} t \Blue{in} p.Tutors.Tutors.Tutors
17.   \}\}
18.\}
\end{Verbatim}
\end{minipage}
\end{tabular}
\end{center}
    \caption{Alloy Model of a Classroom Management System with an oracle and student submission for inv15}
    \label{fig:alloy}
\end{figure*}

The following are equivalent:

\begin{description}
\item [(1)] Complex weighted graph $\mathcal{G}(A)$ is structurally balanced.

\item [(2)] Zero is a eigenvalue of $L$ with eigenvector $\bm{\zeta}=[1\angle\theta_1,\dots,1\angle\theta_N]^T$.

\item [(3)] $D_\zeta:=diag(\bm{\zeta})$ such that $\widehat{A}=D_\zeta^{-1}AD_\zeta$ is nonnegative and $\widehat{A}=[|a_{ij}|]_{N \times N}$.

\item [(4)] $D_\zeta:=diag(\bm{\zeta})$ such that $\widehat{L}=D_\zeta^{-1}LD_\zeta$ has a zero eigenvalue with an eigenvector being $\bm{1}$, where $\widehat{L}=D-\widehat{A}$.
\end{description}

\end{lemma}
\begin{proof}
See proof of Lemma 1 in \cite{sbgcomplete}. 
\end{proof}

Notably, in the notation of \cite{sbg, sbgcomplete}, each column specifies the outer edges of a node, and the row-based adjacency matrix must be transposed before acquiring the Laplacian. 

\subsection{Alloy}

Alloy users write models that describe the properties of the system of interest. Then, the Analyzer helps the user understand their system by displaying the consequences of their properties, helping identify any missing or incorrect properties, and exploring the impact of modifications to those properties.  To achieve this, the Analyzer uses off-the-shelf SAT solvers to search for scenarios, which are assignments to the sets and relations of the model such that all executed formulas hold. If no such scenario can be found, the Analyzer reports that the formulas are unsatisfiable.

To highlight how modeling in Alloy works, Figure~\ref{fig:alloy} depicts the base model of a classroom management system. This model is from the Alloy4Fun dataset, which is a collection of submissions made by novice users learning Alloy~\cite{alloy4fun}. Signature paragraphs introduce named sets and can define relations, which outline relationships between elements of sets. Line 1 introduces a named set \CodeIn{Person} and establishes that each \CodeIn{Person} atom connects to any number of (\CodeIn{set}) \CodeIn{Person} atoms through the \CodeIn{Tutor} relation and each \CodeIn{Person} atom connects to any number of (\CodeIn{set}) \CodeIn{Class} atoms through the \CodeIn{Teaches} relation. Line 5 introduces the named set \CodeIn{Group}, which contains no relations. Line 6 introduces the named set \CodeIn{Class} and states that each class has a set of people assigned to a group using the ternary relational \CodeIn{Groups}. Lines 7 and 8 introduce the named sets \CodeIn{Teacher} and \CodeIn{Student} as subsets (\CodeIn{in}) of \CodeIn{Person}.  

\usetikzlibrary{shapes.geometric, arrows}
\tikzstyle{arrow} = [line width=1.5pt,->,>=stealth]
\tikzstyle{darrow} = [line width=1.5pt,<->,>=stealth]
\tikzstyle{PersonAtom} = [rectangle, minimum width=.7cm, minimum height=.7cm, text centered, draw=black, fill=yellow!75!orange]
\tikzstyle{ClassAtom} = [rectangle, minimum width=.7cm, minimum height=.7cm, text centered, draw=black, fill=orange!75!yellow]
\tikzstyle{State} = [circle, minimum width=0.5cm, minimum height=0.5cm, text centered, draw=black, fill=white]

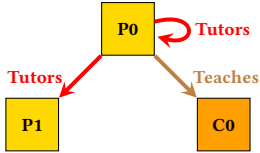
\begin{figure}
\begin{minipage}[t]{.45\columnwidth}
\begin{center}
\footnotesize
\begin{tikzpicture}[baseline,node distance=1.8cm]
\node (P0) [PersonAtom] {\textbf{P0}};
\node (P1) [PersonAtom, below left of=P0] [align=center]{\textbf{P1}};
\node (C0) [ClassAtom, below right of=P0] [align=center]{\textbf{C0}};
\begin{scope}[red]
\draw [arrow] (P0) edge [loop right,line width=1.5pt] node {\textbf{Tutors}} (P0);
\draw [arrow] (P0) -- node[anchor=east] {\textbf{Tutors \ \ }} (P1);
\end{scope}[red]
\begin{scope}[brown]
\draw [arrow] (P0) -- node[anchor=west] {\textbf{ \ Teaches}} (C0);
\end{scope}[brown]
\end{tikzpicture}
\end{center}
\end{minipage}
    \caption{Counterexample highlighting difference between inv15 and inv15oracle}
    \label{fig:ce}
\end{figure}

Predicate paragraphs introduce named first-order, linear temporal logic formulas that can be invoked elsewhere. Figure~\ref{fig:alloy}~(b) depicts the oracle for exercise \CodeIn{inv15}. The predicate \CodeIn{inv15oracle} uses universal quantification (`\CodeIn{all}'),  set multiplicity (`\CodeIn{some}'), set intersection (`\CodeIn{\&}'), transitive closure (`\CodeIn{\^}'), and relational join (`\CodeIn{.}') to express that for every person, there is some intersection between the set Teacher and the set of Tutors reachable from person p. Figure~\ref{fig:alloy}~(c) displays a faulty student submission for this exercise.  The student attempts to use nested quantification to explicitly outline that a teacher should be reachable in 1 to 3 traversals down the Tutor's relation. The student submission assumes incorrectly that a person's tutor relation captures the set of all people who tutor $p$, but the tutor relation captures who $p$ themselves tutors. Therefore, the student submission can be corrected by inserting the transpose operator, e.g., ``\CodeIn{p.\ATilde{Tutors}}.'' 

Figure~\ref{fig:ce} displays one of the counterexamples produced when the student submission is checked against the backend oracle in Figure~\ref{fig:alloy}~(b). The depicted scenario is valid for the oracle solution but not for student submission. To find this counterexample, an Alloy command is run that invokes the backend SAT solver. All commands require a scope, which places an upper bound of the universe of discourse. The default scope is 3, which means Alloy tries to find a counterexample highlighting that the two formulas are not equivalent using up to 3 Person, 3 Group, and 3 Class atoms. 


\subsection{The Alloy Grammar}
\begin{figure*}[!htbp]
\centering
\includegraphics[width=\textwidth,height=\textheight,keepaspectratio]{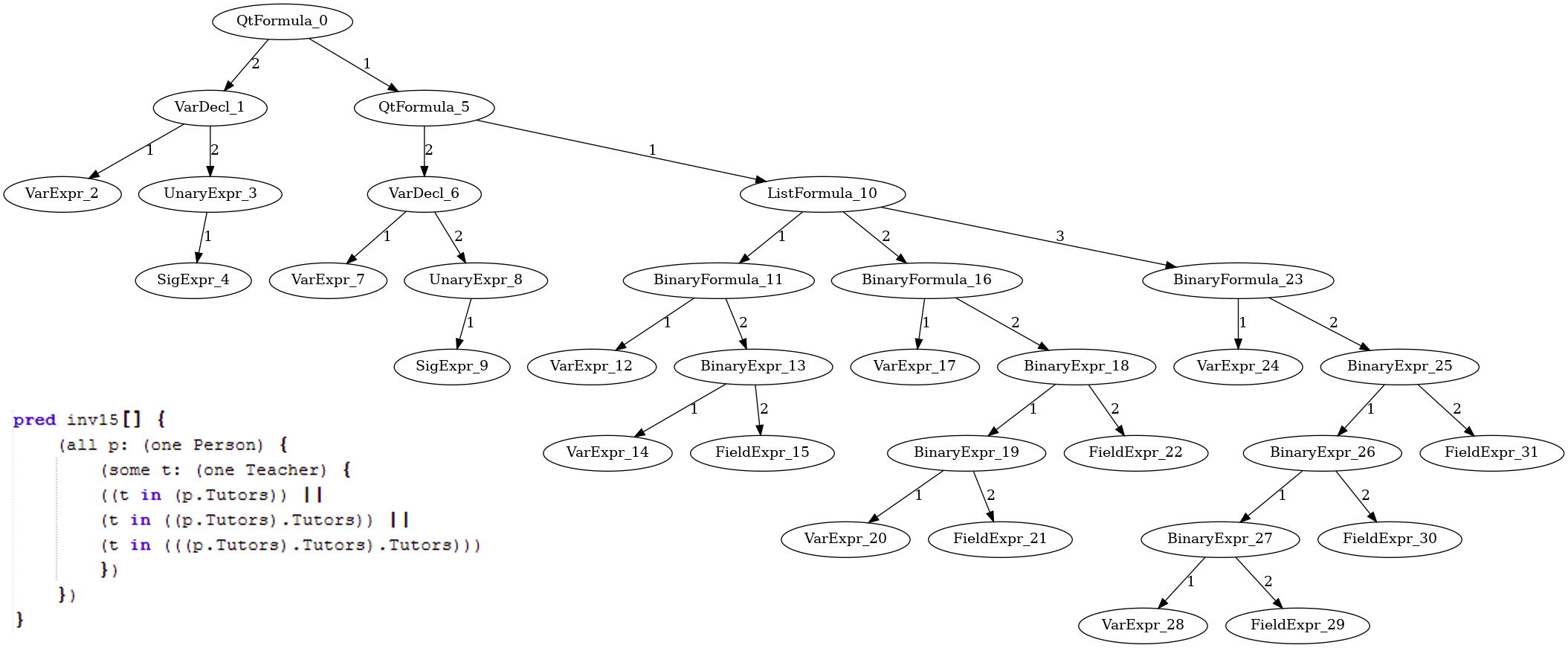}
\caption{An instance of an Alloy predicate in the dataset parsed with pretty string and its raw AST representation. }
\label{fig:fig1}
\end{figure*}

We first highlight aspects of the Alloy grammar \cite{alloy} that influence how we construct the graph representation. 
An Alloy model consists of a set of declarations with signatures as basic types, functions as processes, and predicates as logical arguments. Our dataset Alloy4Fun \cite{alloy4fun} covers only simple predicates without references to external libraries; therefore, creating representations for those code segments included in the dataset is an ideal starting point. Figure~\ref{fig:fig1} gives an example of an Alloy predicate in the dataset that contains 32 AST nodes. From this representation, it is obvious that the raw AST representation is relatively huge and complicated, even with a short segment of code enveloped in a single predicate, mostly due to the redundancy of semantically identical nodes that appear multiple times in the AST since they appear multiple times in the original code. In the given example, there are three subset operators ``in'' and 6 relational join ``.'' operators that are all treated as distinct nodes in the raw AST representation. 

A predicate defines a scope that contains an expression while itself is a solid, invariant root for any fixing pair. We can easily enumerate all possible nodes under an expression (or formula): the finitely enumerable types of expressions or formulae, plus variable declarations that only occur under the quantifiers of first-order logic. In this paper, we assign them fixed signatures, yet an adaptive signature system could be used in the case of an extension of the CSBASG into either more types of nodes or extra optimizations.

Another property of Alloy is that almost all expression or formula nodes have a finite and fixed number of children, for instance, 1 for unary operators, 2 for binary operators, and 3 for the if-then-else sentences. In these cases, there could be a trivial, polynomial-based implementation of the $|w_{ij}|$ values for the edges between such a node $v_i$ and one of its children $v_j$. Even so, there are some exceptions, such as a function call (having an indefinite number of parameters), a list expression or formula (for example, the consecutive logical ANDs and ORs for a conjunctive or disjunctive normal form), or a set of variable declarations for the same quantifier. We will handle them on a case-by-case basis in Section 3. 
\section{CSBASG Construction}
Intuitively, to create an ASG following \textit{Definition 3}, we first need to do a pre-order traversal on the AST 
to identify and combine the semantically identical nodes. We construct an isomorphism between $\mathcal V$ and $N\cup\Sigma$, then rewrite each AST link into an ASG edge. The isomorphism is trivial since AST nodes with the same word are combined into the same node in the ASG. Nonetheless, constructing the ASG edge set by the AST relations $\xi$ requires a cautious step: merging two AST nodes requires clarifying the orders of the visit to avoid confusion. 
For instance, in the example given in Figure 1, if we consider the binary operator ``IN'' as a single node without specifying the order of execution, we would not specify which of the following would be put as the right-hand operand: ``\CodeIn{p.Tutors}'', ``\CodeIn{p.Tutors.Tutors}'' or ``\CodeIn{p.Tutors.Tutors.Tutors}''.   A naive approach that deviates a bit from the ASG definition is to use two numbers for the weight values $w_{ij}$ instead of one, that is, for any of the existing link $x_2\in \xi(x_1)$ in the AST, defining $v_1=\sigma(x_1), v_2=\sigma(x_2)$. Then, we create an edge between $v_1, v_2$ with a weight-pair $(\omega, t)\in \mathbb N \times \mathbb N$, specifying that $x_2$ is the $\omega$-th child of $x_1$ in the order defined for $\xi(x_1)$, and it is the $t$-th time to visit $v_1$ in the program logic. 

Nevertheless, we expect that for a matrix form of the ASG, we could take advantage of their linearity, especially for the potential application to machine learning operations. In addition, a matrix form of the graphs could be more easily compared with respect to multiple implementations of a predicate or other code segments of interest. With the naive construction above, it is challenging to proceed since the mapping from the pair given above to the edge value that serves as an entry in the adjacency matrix is nontrivial. Here, we could define $\mathcal M:\mathbb N\times \mathbb N\to\mathbb R$ as mapping such that $\mathcal M(\omega, k)$ gives the entry in the matrix form of an ASG if it is the sole AST link in the ASG. The mapping could be defined arbitrarily, provided it maintains injectivity to avoid the simple, different-edge-same-value confusion (we will call it Type 1 Confusion later). Such forms the real part (magnitude) of the edges in the CSBASG. 

Considering all the above, we formally define the complex-valued structurally balanced ASG as a subset of the general ASG constructions that maintain the structural balance. 
\begin{definition}[Complex-valued Structurally Balanced ASG (CSBASG)]
A Complex-valued Structurally Balanced ASG is a type of ASG $\mathcal G\equiv (G, \mathcal V, \mathcal E, r)$ where  there exists a vector $\{\theta_1, \theta_2, ..., \theta_{|\mathcal V|}\}$, such that the weights in each of the edge $(v_i, v_j, w_{ij})\in\mathcal E$ satisfy $w_{ij}=|w_{ij}|\angle (\theta_i-\theta_j)$. For each of the node $v_i\in\mathcal V$, $\theta_i\in (-\pi, \pi)$ is called the signature of $v_i$. 
\end{definition}

\textit{Algorithm 1} gives the high-level process to convert an AST to a CSBASG. It follows the process of an abstract interpretation, where $\vec{t}$ in the recursive part gives a simplified interpretation state that tracks the times of node visitations in the pre-order traversal of the AST. We first initialize the nodemap and the list of angular signatures for each node, with the only element being the dummy overall root. Then, we assign each node an angular signature with regard to a predefined assignment, which outlines every unique signature. The number of unique angular signatures (semantically distinct nodes) is the number of rows and columns of the resulting adjacency matrix. After that, we initialize a state vector $\vec{t}$ to count the times of the order of visitation to each node. A recursive function then iterates over the AST beginning from the overall root, updates the adjacency matrix for each downlink from the root node according to the composition function $\beta$ and the encoding method $\mathcal M$, while updating the counter of the given node with the corresponding entry of the state vector signaling the pre-order time of identical appearance in the AST before recursively visiting the subtree with the new root being a child of the local root. 

In general, with a static signature assignment $\varphi$, the algorithm terminates in $O(|X|)$ since each of the nodes in the AST is visited exactly once and $|\mathcal V|<|X|$ strictly. We say an ASG is \textit{generated} by an AST if the AST was converted to the same ASG with the same signature assignment and magnitude mapping. 

\begin{algorithm}[t]
\begin{algorithmic}[1]
\caption{Convert AST to CSBASG}
    \State{\textbf{Input:} an AST $T\equiv (G,X,r,\xi,\sigma)$ parsed from the original code of an Alloy predicate, with root node $r$; 
    an injective function $\varphi: N \cup \Sigma\to (-\pi, \pi)$ to map the words to their unique signature, and another injective function $\mathcal M:\mathbb N\times \mathbb N\to \mathbb R$ outputs the magnitude of an AST edge given the positional argument of a link and the time has the sourcing node been visited. For the entries, we use $\beta$ as a recursive encoder that calculates the matrix entries iteratively. }
    \State\textbf{Output:} a CSBASG $\mathcal G(A)$ representing $T$ with its adjacency matrix $A$, and the signatures of the nodes $\{\theta_0, \theta_1, \theta_2, ..., \theta_{|\mathcal V|}\}$, each corresponds to a column of $A$. 
    \State $\vec{\theta} \gets [0]$
    \State $nodemap \gets [(r, 0)]$
    \State $\mathcal V \gets \{r\}$
    \For {$i \gets 2, ..., |X|$}
        \State $node \gets X_i$
        \State $\theta_i \gets \varphi(\sigma(node))$
        \If {$\theta_i \notin \vec{\theta}$}
            \State {$\vec{\theta}\gets \vec{\theta} :: \theta_i$}
            \State {$\mathcal V\gets \mathcal V :: node$}
        \EndIf
        \State {$nodemap \gets node map :: (node, \theta_i)$}
    \EndFor
    \State $A_{\vec{0}}\gets [0]_{|\mathcal V|\times |\mathcal V|}$
    \State $A\gets$ \Call{encode-recursive}{$r, \vec{0}_{|\mathcal V|}, A_{\vec{0}}$}
    \State\Return $A, \vec{\theta}$
    \Function {encode-recursive}{$r_t, \vec{t}, A_{\vec{t}}$} \
        \State $\theta_t \gets nodemap.\Call{get}{r_t}$
        \State $i \gets \vec{\theta}.\Call{get-index}{\theta_t}$
        \State $t \gets \vec{t}.\Call{get}{i}$
        \State $\vec{t} ' \gets \vec{t}$
        \State $\vec{t} '[i] \gets t + 1$
        \State $A_{\vec{t}'}\gets A_{\vec{t}}$
        \For {$\omega \gets 1, ..., |\xi(r_t)|$}
            \State $c\gets \xi(r_t).\Call{get}{\omega}$
            \State $\theta_{c} \gets nodemap.\Call{get}{c}$
            \State $j \gets \vec{\theta}.\Call{get-index}{\theta_{c}}$
            \State $A_{\vec{t}'}[j, i]\gets \beta(\mathcal M(\omega, t), A_{\vec{t}}[j, i])\cdot \angle (\theta_i - \theta_j)$
            \State $A_{\vec{t}'}\gets \Call{encode-recursive}{c, \vec{t}', A_{\vec{t}'}}$
        \EndFor
        \State\Return $A_{\vec{t}'}$
    \EndFunction
\end{algorithmic}
\end{algorithm}

By \textit{Definition 5}, for a given Alloy predicate as an input, we define a fixed unique root $r\equiv v_1$, which is the predicate itself, and such root node is unique in each predicate. We can intuitively assign $\theta_1 = \varphi(\sigma(r)) \equiv 0$ as a starting point. Besides, we expect the same input parameters for the pairs of predicates to be directly compared, so the overall root node of each predicate only has one child, the local root expression or formula. Therefore, the only work left is to define the signature assignment function $\varphi$ on the subset that could be present under a tree rooted by an expression or formula.

The function,  $\mathcal M$, could be defined in multiple ways if it maintains a one-on-one correspondence between the syntactic positions and the execution or tree-walking order. However, for an AST with multiple syntactically and semantically identical nodes, there can be multiple edges that connect them, but those edges could have different positions $\omega$ and thus different $\mathcal M$ values. So, given an AST, we can define the three types of confusion that could happen in an ASG construction: 
\begin{definition}[Completeness of CSBASG]
    Given a class of AST $T\equiv (G, X, r, \xi, \sigma)$, we say a CSBASG construction is \textbf{complete} if the parser $L$ outputs the AST correctly by the generated CSBASG in Algorithm 1; that is, none of the confusions formed in the construction of $\mathcal M$ and $\beta$ mappings: 
    \begin{itemize}[leftmargin=.4cm]
        \item If $\mathcal M$ is not injective, i.e., for two edges with positional values $\omega_1\neq \omega_2$ or tree-walking orders $t_1\neq t_2$, $\mathcal M(\omega_1, t_1)=\mathcal M(\omega_2, t_2)$. This situation is a Type 1 Confusion, or Incomplete Edge Mapping Confusion. 
        \item If an ASG parser $L$ cannot parse a multi-edge in an ASG that corresponds to two AST links that have the same parent and different but syntactically and semantically equal children with different positions, e.g., $(a+b)-(c+d)$ where the central binary operator $-$ has two syntactically and semantically equal children of binary operator nodes $+$, then the situation causes a Type 2 Confusion or Twin Children Confusion. 
        \item If an ASG parser $L$ cannot parse a multi-edge in an ASG that corresponds to two AST links connecting semantically and syntactically equivalent pairs of nodes, but the parent nodes are at different positions in the AST, e.g., $(a\land b) \lor (a\land c)$, the ASG node $``\land"$ has two left children of terminal node $a$, then the situation causes a Type 3 Confusion or Twin Parent Confusion. 
    \end{itemize}
\end{definition}

Here, the goal becomes creating the mappings $\mathcal M$ and $\beta$ that ensure the parser $L$ is free of confusion. 
\subsection*{Polynomial-based Static CSBASG Encoding}
By \textit{Definition 6}, intuitively, the signature assignment $\varphi$ is not relevant to the completeness of a CSBASG; that function could be arbitrarily defined or even \textit{learned}. However, the processes directly related to the fidelity of the representation scheme must be pre-defined to ensure the completeness of the CSBASG. A naive implementation could consider the $\omega$ positional values as sequential numbers $1, 2, ...$ and add them together for the representation; however, if there is a ternary operator $tor$, then $tor(a, a, c)$ will have its children-position map be defined as $\{a:3, c:3\}$, causes confusion that has two children with a position 3 (and therefore, missing positions 1 and 2), causes a Type 2 confusion. 

Here, we introduce a polynomial-based, integer-magnitude static CSBASG encoding for a theoretical language: each AST generated from its CFG has a maximum number of children $p$. Then, let $T$ be the maximum number of syntactic and semantically identical nodes in the code segment. For the AST link drawn from a node to its $(\omega + 1)$-th child, let $\mathcal M(\omega, t)=2^{(t-1)p+\omega}$ and $\beta$ be a simple summation of the $\mathcal M$ values. Our construction of the scheme was defined in the order of $O(2^{pT})$. It could be despairing at a glance, yet consider that in many languages without a list expression, $p\leq 5$ could be achieved, and $T$ only increases with the density and size of the code chunk, and raw AST representations also suffers such a polynomial growth. 
\begin{lemma}
The polynomial-based CSBASG magnitude encoding is complete and optimal; there is no complete encoding with the maximum magnitude of the ASG matrix entry less than $O(2^{pT})$. 
\end{lemma}
\begin{proof}
Completeness follows since $1\leq\omega\leq p$ always holds, for each integer $k\in\mathbb Z$ such that $\mathcal M(\omega, t)=2^k$, there is a unique pair of $\omega$ and $t$, there is no Type 1 Confusion. Consider multiple AST links from the syntactically and semantically same parent node to the syntactically and semantically same child, each with $\omega_1, \omega_2, ..., \omega_n$ as their positions, at $t_1, t_2, ..., t_n $-th visit of the parent node. Then, $\sum_{j=1}^n 2^{(t_j-1)p+\omega_j}=\sum_{k=0}^{pT}\mathbb{I}\{\exists{j=1,...,n}|k=(t_j-1)p+\omega_j\}2^k$ is a unique integer representation of the magnitude of an adjacency matrix entry in the ASG, so it is also free of Type 2 or Type 3 Confusion, gives the completeness of the representation. Optimality follows since each positive integer has a binary form $k=\sum_{j=0}^\infty c_j2^j$, $c_j\in\{0,1\}$, which each $c_j=1$ corresponds to an edge with $j=(t-1)p+\omega$. 
\end{proof}

By \textit{Lemma 2}, the summation-of-exponentials gives the most compact and accurate solution for the complete enumeration of the graph, and the creation of such a graph also depends on the enumerability of the grammar. However, a representation within the 32- or 64-bit bound could be achieved for a less-scaled language like Alloy. A heuristic generated by representation learning could also be achieved by learning the logarithmic parameters for a more general-purpose language with more syntactic elements and a larger AST. 


\subsection*{The Decoding Algorithm}
The decoding process is straightforward in theory. Words of the code are stored in correspondence to rows and columns of the matrix; each row or column corresponds to an individual word, and vice versa. Formally, we designate $L: \mathcal G\to \mathcal T$ as the parser that translates an ASG back to its uniquely corresponding AST. $L$ is therefore a counterpart of \textit{Algorithm 1} and bounded by the constructions of $\mathcal M$ and $\beta$ functions. For each of the nonzero entries in the ASG adjacency matrix, we can break it into concrete links by reversing $\mathcal M$ and $\beta$. For the binary polynomial-based magnitude encoding given above, we take the integer value of the magnitude $|a_{ij}|$ and break it with regard to its binary form: each digit signing $1$ in the binary form corresponds to a concrete AST link. In this case, an ASG node is broken into a set of AST nodes, each with its corresponding outer edges by the visit order. Since the possible edges are all examined once and the size of the entries determines the binary forms, we have a complexity of $O(|\mathcal E|\log(\max(w_{i,j})))$ or $O(|\mathcal V|^2\log(\max(w_{i,j})))$ for the decoding process. 



\section{Application of CSBASG on Alloy}

\begin{figure}
\centering
\includegraphics[width={0.45\textwidth} ,]{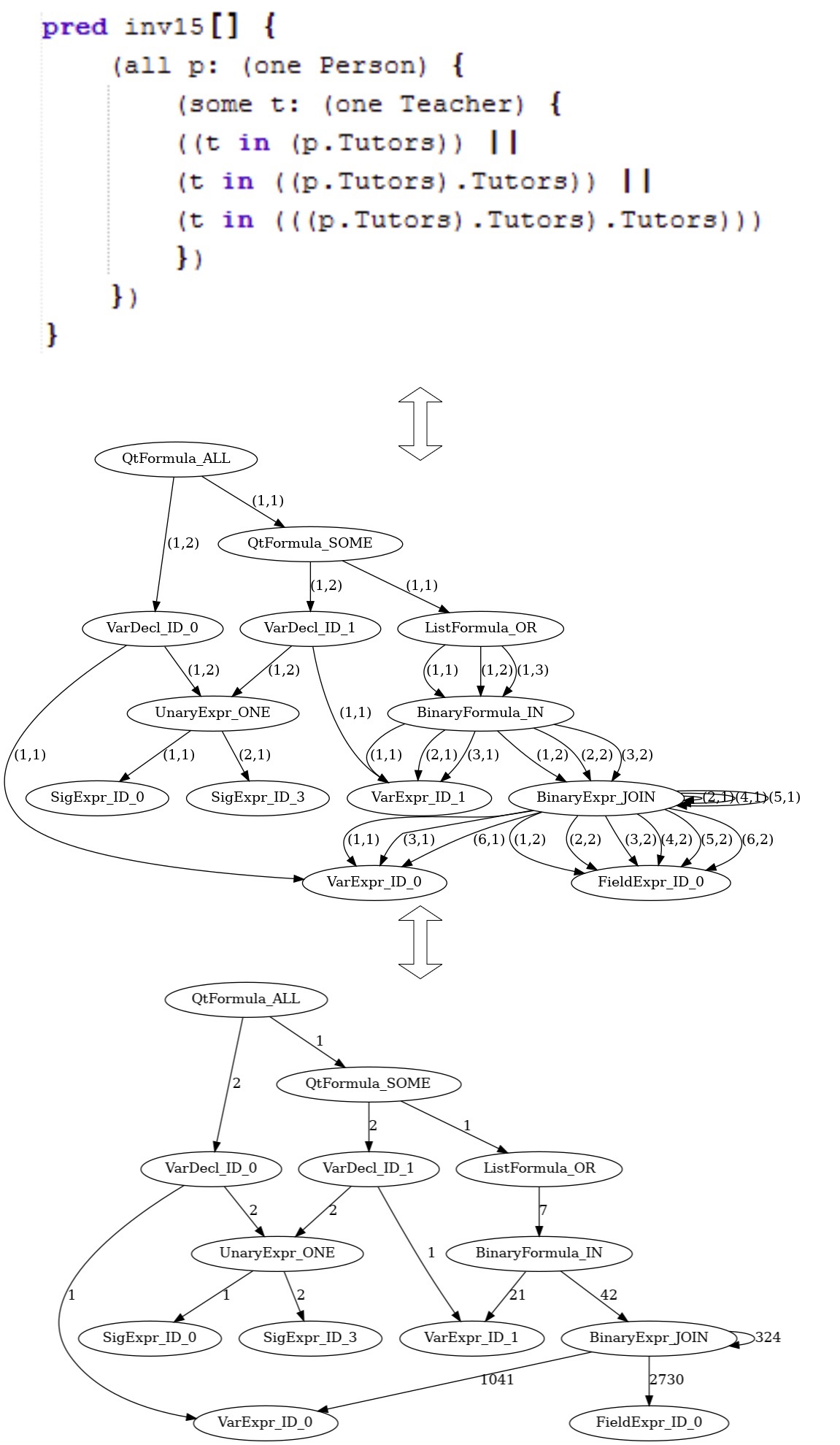}
\caption{An example illustrating the compactness and simplicity of CSBASG; the original code (upper) could be expressed as an equivalent multigraph (middle) with the AST in \textit{Figure 1}, and turns into the adjacency relationship (lower) with the polynomial-based encoding. With the same predicate in Figure 1, the number of nodes reduced significantly from 31 to 12. All edges are now assigned an integer weight value, which forms an adjacency matrix. }
\label{fig:fig2}
\end{figure}

There are two main features of Alloy that make for an ideal starting point to apply our CSBASG representation. 
First, as a first-order logic-based declarative language, we can easily construct a finite set of symbols given a model's source code. Second, while there are a few infinitely extensible structures in its grammar, which does limit the usage of empirical estimates for the maximum nodes under a given structure, even the infinitely extensible structures could be assumed to have a relatively low count of children. 

To begin, we utilize a parser that classifies the syntactic properties of the code keywords into a limited number of categories. After that, we need to set a $p$ value corresponding to the maximum number of children for each category, for instance, $1$ for unary operators, $2$ for binary operators, $3$ for if-then-else, and a more significant number for the several uncovered infinite categories. This could cause ambiguity in some edge cases. Still, since Alloy is used to model structures within a bounded scope for each category of sets, there will unlikely be any long enumerations over 10 or 15 items, so we use $p=17$ for the categories without a precise, finite maximum children number. While completeness is not guaranteed, we aim to solve completeness for the majority of cases. While it is possible to break the enumerations by putting the list of items into a linked structure, in most cases, the tradeoff of completeness in the edge cases delivers more benefits in the intuitiveness and compactness of the representation. 

Inside each category, the nodes are differentiated by their semantic contents, which are still finitely enumerable since we have access to the signatures defined within a model file, and there are almost no numerical constants. Moreover, the numerical constant will be bound by the scope. Therefore, it is safe to assume that constants are limited to booleans and small integers. In our approaches, we divided the AST nodes under a paragraph of an expression or formula into 19 categories, plus the zero-signature node capturing the local root. Devolved from the syntactic categories are semantic subcategories, and the number of subcategories within a syntactic category ranges from 1 for the simple, invariant nodes to 32, which is the number of possible binary expressions. 

Finally, we need to assign the angular signatures for each node. In our practice, we used a simple encoding by dividing the interval of a circle $(-\pi, \pi)$ into equal-length intervals with $D = \pi / 13$ as the length of each interval. For each interval, we assign subticks that divide it into 128 or 65536 sectors, depending on the syntactic category. By the scheme above, we have obtained each AST node's syntactic-semantic pair $(\Psi, \psi)$ using the results generated from the parser. Then, we map them on the unit circle as their signatures: each syntactic category goes with the beginning of a $\pi/13$-length interval, and each subtick above the interval starting point corresponds to a semantic subcategory. Note that this is just a demonstration of encoding and is only defined for convenience but not learned for optimality, and the angular correspondence could be any bijective mappings.

Figure~\ref{fig:fig2} gives an example of the encoding from an Alloy predicate AST to its corresponding CSBASG using a polynomial-based static scheme. The syntactic-semantic pairs in the table give the properties of each node in the AST and combine the AST nodes with identical semantic and syntactic properties. The angular signature assigned to each node is $\Psi D+\psi\delta$, where $\delta=D/128$ for expression or formula nodes and $\delta=D/65536$ for the relational declarations that would appear below a quantifying expression or formula, since by Alloy syntax multiple variables could be declared under the same relational declaration node. {In our example, there are six JOIN binary operators (.), which are represented as six distinct nodes in the original AST; in the ASG approach, the nodes are combined as a single node with the label BinaryExpr.JOIN. }

The complex values could be used to compare two ASGs with shared parts, which will be discussed in Section 5.3. 


\section{Experiment}

We address the following research questions:

\begin{itemize} \itemsep=0em
    \item \textbf{RQ1:} What degree of compactness does the CSBASG representation of an Alloy model achieve? 
    \item \textbf{RQ2:} Does our CSBASG representation of an Alloy model enable comparison between different predicates? 
\end{itemize}

\begin{table}[]
\caption{Overview of problem set complexity.}
\label{tab:model_states}
\begin{tabular}{lrrrr}
\toprule
    \textbf{Problem} & \multicolumn{1}{c}{\textbf{\#Sig}} & \multicolumn{1}{c}{\textbf{\#Rel}} & \multicolumn{1}{c}{\textbf{\#Exe}} & \multicolumn{1}{c}{\textbf{\#AST}} \\
\midrule
classroom\_fol & 5 & 3 & 15 & 10.00 \\
classroom\_rl & 5 & 3 & 15 & 10.00 \\
cv\_v1 & 5 & 4 & 4 & 19.75 \\
cv\_v2 & 5 & 4 & 4 & 21.75 \\
lts & 3 & 1 & 7 & 19.71 \\
production & 5 & 3 & 4 & 14.25 \\
train & 6 & 3 & 18 & 23.44 \\
trash\_rl & 3 & 1 & 10 & 4.80\\
\bottomrule
\end{tabular}

\end{table}

\subsection{Dataset}
Our dataset encompasses 6,307 models from Alloy4Fun\cite{alloy4fun}, sourced from 8 different problem sets, filtered from a larger dataset to ensure each is compliable, runnable, nonempty, and the student-written model is not completely identical to the ground truth. For each problem set, Alloy4Fun has a base model that outlines all the signatures and relations, as well as a collection of empty predicates with English descriptions. Users can then attempt to fill in the predicate to match the English description, and their submission is checked against a backend oracle for correctness. Table~\ref{tab:model_states} gives an overview of the complexity of each problem set: Column \textbf{\#Sig} is the number of signatures,  \textbf{\#Rel} is the number of relations, \textbf{\#Exe} is the number of different predicates uses can try to complete on Alloy4Fun, and \textbf{\#AST} is the average number of AST nodes in the oracle solutions. 

Our 6,307 models contain the base signatures from the Alloy4Fun problem set along with two predicates: a student-written predicate named InvX (or PropX, X is an identifying number) and another predicate called InvXC that is oracle solutions from Alloy4Fun. For each model, we run two predicates, \textit{overconstrained} and \textit{underconstrained}, which determine if there are cases that satisfy InvX but not InvXC, or vice versa. If a model has no cases for both \textit{overconstrained} and \textit{underconstrained} predicates, then InvX is formally correct. This gives four categories of the models: correct ($InvX=InvXC$), overconstrained only ($InvX\subsetneq InvXC$), underconstrained ($InvXC\subsetneq InvX$), and both overconstrained and underconstrained ($InvX\not\subset InvXC \land InvXC\not\subset InvX$), covers all possible cases. 

\begin{table}
\centering
\caption{Compactness test results of the dataset.}
\label{tab:compact}
\begin{tabular}{lrlrr}
\toprule
    \multicolumn{1}{c}{\textbf{Problem}} &  \multicolumn{1}{c}{\textbf{\#Models}} & \multicolumn{1}{c}{\textbf{Mut.Type}} &  \multicolumn{1}{c}{\textbf{Mut.\%}} & \multicolumn{1}{c}{\textbf{Oracle\%}} \\
\midrule
lts & 113 & OVER & 24.65 & 40.52 \\
lts & 138 & BOTH & 24.76 & 35.9 \\
lts & 66 & UNDER & 27.14 & 34.31 \\
lts & 61 & CORRECT & 36.66 & 39.19 \\
trash\_rl & 150 & OVER & 13.42 & 7.10 \\
trash\_rl & 267 & BOTH & 14.85 & 6.99 \\
trash\_rl & 62 & UNDER & 31.67 & 11.68 \\
trash\_rl & 252 & CORRECT & 20.25 & 6.73 \\
train & 170 & OVER & 23.47 & 25.23 \\
train & 277 & BOTH & 28.18 & 28.54 \\
train & 174 & UNDER & 32.70 & 24.43 \\
train & 75 & CORRECT & 29.44 & 19.43 \\
classroom\_fol & 223 & OVER & 30.48 & 16.86 \\
classroom\_fol & 1115 & BOTH & 36.48 & 26.08 \\
classroom\_fol & 166 & UNDER & 37.19 & 28.54 \\
classroom\_fol & 495 & CORRECT & 32.83 & 14.67 \\
cv\_v2 & 40 & OVER & 38.28 & 43.79 \\
cv\_v2 & 21 & BOTH & 38.37 & 43.08 \\
cv\_v2 & 12 & UNDER & 29.16 & 36.21 \\
cv\_v2 & 31 & CORRECT & 30.95 & 40.47 \\
cv\_v1 & 118 & OVER & 36.05 & 41.48 \\
cv\_v1 & 68 & BOTH & 32.49 & 39.97 \\
cv\_v1 & 54 & UNDER & 32.03 & 37.94 \\
cv\_v1 & 52 & CORRECT & 31.73 & 39.83 \\
classroom\_rl & 223 & OVER & 30.48 & 16.86 \\
classroom\_rl & 1115 & BOTH & 36.48 & 26.08 \\
classroom\_rl & 166 & UNDER & 37.19 & 28.54 \\
classroom\_rl & 495 & CORRECT & 32.83 & 14.67 \\
production & 25 & OVER & 18.84 & 20.73 \\
production & 22 & BOTH & 16.02 & 20.41 \\
production & 36 & UNDER & 22.89 & 22.80 \\
production & 25 & CORRECT & 14.75 & 21.12 \\
\bottomrule
\end{tabular}
\end{table}

\subsection{Metric 1: Compactness of Representation}
We evaluate the compactness of the representation by comparing the number of nodes in the ASG representation compared to the number of nodes in the AST. Table~\ref{tab:compact} gives an overview of these results. Column \textbf{Problem} displays the problem set from  Alloy4Fun~\cite{alloy4fun}, and column \textbf{\#Models} displays the number of unique student submissions. From each problem set, the student-written submissions, which we refer to as mutant predicates, are categorized further into four mutant types given their coverage of the set indicated by the ground truth: CORRECT, OVERconstrained, UNDERconstrained, or BOTH over- and underconstrained (Column \textbf{Mut.Type}). The percentages given are the ratio of the reduction of nodes in the ASG representation compared to the raw AST for the student submission (Column \textbf{Mut.\%}) and the oracle solution 
(Column \textbf{Oracle\%}), e.g., there are 32.83\% fewer nodes in the ASG of correct classroom\_fol student submissions than the AST. 

Overall, we achieved an impressive 27.25\% reduction of nodes over the dataset with no loss of information. Interestingly, a smaller oracle does not guarantee less reduction. While  the model with the smallest oracles by number of AST nodes, trash\_rl, does in fact see the smallest reduction, the train model has the largest oracle but three models have larger reductions in nodes. All told the reduction the CSBASG provides over the AST is tied to the likelihood that the formula to be expressed has redundancy within its structure.  
Note that since the dataset is mostly single predicates designed as after-class practices for college students, the probability of semantic identical nodes in the AST is relatively low, and for a sub-AST with a large scale, the performance would increase accordingly in the single-metric of node reduction. Moreover, the size of the adjacency matrix scales up with $O(N^2)$ with the number of nodes so that a denser matrix could be achieved with the ASG scheme.

\begin{figure*}
\centering
\includegraphics[width=\textwidth,height=\textheight,keepaspectratio]{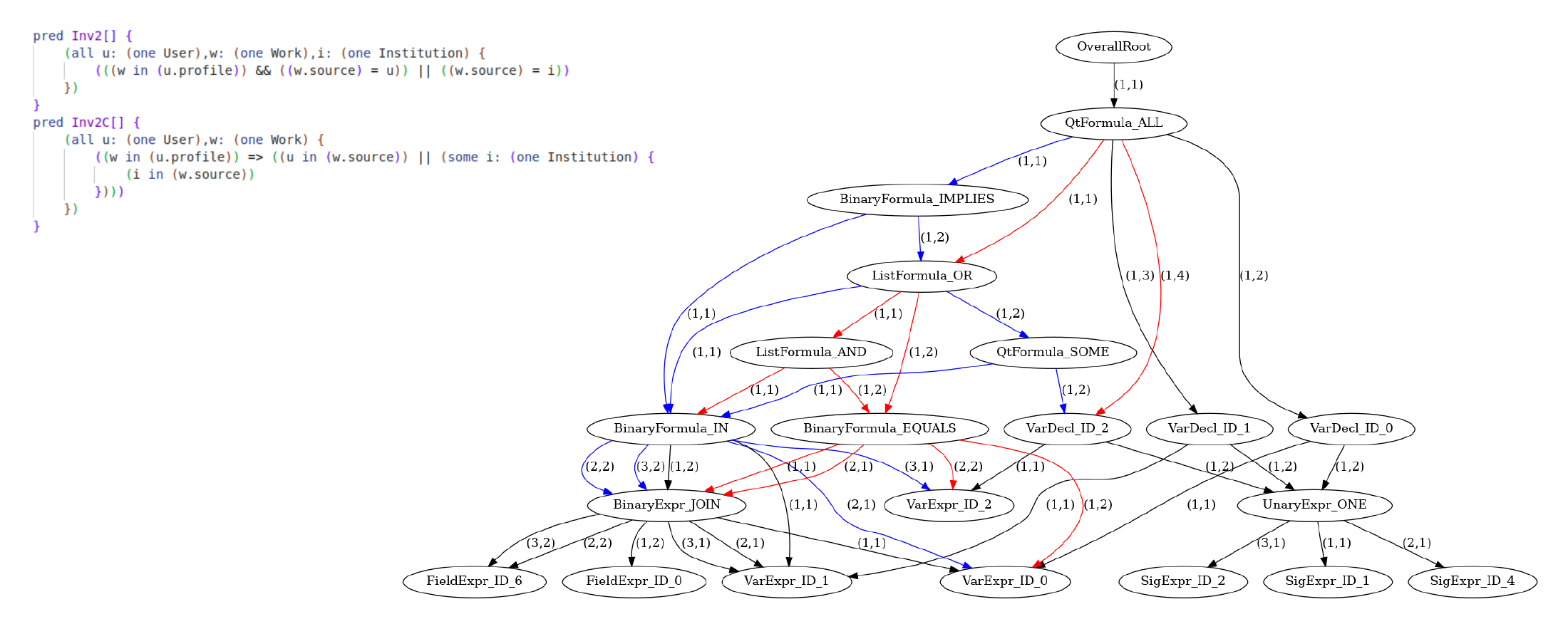}
\caption{An example of a pair of student-written mutant predicate (Inv2) and the ground truth (Inv2C). The ASG edges in form $(x,y)$ indicate the $x$-th visit of the sourcing node, and the positional relationship between the nodes is $y$. Black edges are present in both predicates, red edges are present in the mutant only, and blue edges are present in the ground truth only.
}
\label{fig:compare}
\end{figure*}

However, this metric should not be confused with saving spaces. Since in an AST the entries in the adjacency matrix are either 0 (indicating no links) or 1 (indicating a direct link), which only occupy 1-bit per entry in an optimized language or computational method, but \textit{Lemma 2} indicates that there is no complete encoding without information loss that takes a number smaller than $O(2^{pT})$ which occupies $pT$-bits at least, so space saving is neither guaranteed nor indicated by the inferences and experiment above. 

\begin{table}
\centering
\caption{Comparability quantification results of the dataset.}
\label{tab:compare}
\begin{tabular}{lrlrr}
\toprule
    \multicolumn{1}{c}{\textbf{Problem}} &  \multicolumn{1}{c}{\textbf{\#Models}} & \multicolumn{1}{c}{\textbf{Mut.Type}} &  \multicolumn{1}{c}{\textbf{Mut.\%}} & \multicolumn{1}{c}{\textbf{Oracle\%}} \\
\midrule
lts & 113 & OVER & 69.66 & 31.03 \\
lts & 138 & BOTH & 71.46 & 36.55 \\
lts & 66 & UNDER & 60.87 & 42.91 \\
lts & 61 & CORRECT & 51.61 & 49.31 \\
trash\_rl & 150 & OVER & 84.08 & 61.93 \\
trash\_rl & 267 & BOTH & 89.39 & 62.35 \\
trash\_rl & 62 & UNDER & 90.21 & 75.24 \\
trash\_rl & 252 & CORRECT & 90.64 & 73.49 \\
train & 170 & OVER & 70.06 & 49.00 \\
train & 277 & BOTH & 75.16 & 50.97 \\
train & 174 & UNDER & 73.29 & 60.55 \\
train & 75 & CORRECT & 66.66 & 63.86 \\
classroom\_fol & 223 & OVER & 77.54 & 63.51 \\
classroom\_fol & 1115 & BOTH & 79.21 & 64.01 \\
classroom\_fol & 166 & UNDER & 69.61 & 61.49 \\
classroom\_fol & 495 & CORRECT & 82.07 & 71.77 \\
cv\_v2 & 40 & OVER & 65.78 & 33.73 \\
cv\_v2 & 21 & BOTH & 67.67 & 35.04 \\
cv\_v2 & 12 & UNDER & 70.70 & 36.82 \\
cv\_v2 & 31 & CORRECT & 63.93 & 38.38 \\
cv\_v1 & 118 & OVER & 58.99 & 34.84 \\
cv\_v1 & 68 & BOTH & 66.96 & 36.03 \\
cv\_v1 & 54 & UNDER & 69.96 & 37.52 \\
cv\_v1 & 52 & CORRECT & 62.88 & 38.86 \\
classroom\_rl & 223 & OVER & 77.54 & 63.51 \\
classroom\_rl & 1115 & BOTH & 79.21 & 64.01 \\
classroom\_rl & 166 & UNDER & 69.61 & 61.49 \\
classroom\_rl & 495 & CORRECT & 82.07 & 71.77 \\
production & 25 & OVER & 71.80 & 45.60 \\
production & 22 & BOTH & 70.67 & 42.82 \\
production & 36 & UNDER & 43.86 & 45.62 \\
production & 25 & CORRECT & 68.12 & 49.4 \\
\bottomrule
\end{tabular}
\end{table}

\subsection{Metric 2: Comparability of Representation}
Comparing a pair of code snippets to train a model for automated correction could be tedious, especially for syntactically correct but faulty predicates. Therefore, the CSBASG representation's comparability could enable graph mutations to annotate the pairs and create training data. We will call an atomic mutation either adding or removing an edge in the ASG. A relationship change between two nodes could also be represented as removing the old edge and adding the new one.  To explore this idea, we have implemented a simple formula to decode the multiedges created when constructing the CSBASG using the polynomial method. 


To get a holistic view of the comparability after decoding the CSBASG, we count the percentage of the shared edges among the predicate pairs. Table~\ref{tab:compare} gives an illustration of the per-category performance with regard to this metric. Again we present the problem set (Column \textbf{Dataset}), the number of submissions (Column \textbf{\#Models}), and the mutant type of the submission (Column \textbf{Mut.Type}). Then, the next two columns present the percentage of shared edges. Column \textbf{Mut.\%} is the average percentage of edges the mutant student submissions share with the oracle, and Column \textbf{Oracle\%} is the average percentage of edges the oracle solution shares with the mutant student submission. The results are promising and show significant evidence of comparability: on average, 60.74$\%$ of edges in the oracle solution also appear in the student-written counterpart, and on average, 77.37$\%$ of edges in the student-written code also appear in the oracle solution. 

To take a closer look at comparability, Figure~\ref{fig:compare} gives an example of the decoded CSBASG construction for comparing a pair of predicates in the same declarative environment, which, in our case, is the faulty and correct solution to the same problem. The common parts, depicted as black edges, show the correct declaration of the first two variables by the student given the right part of the graph, and the trivial correctness of the fields associated with the variables as elements of the pre-declared sets are shown in the bottom-left corner. However, most of the graph shows significant deviation of logic, as expected since the student predicate is \textit{both} overconstrained and underconstrained. Straightforwardly, each of the red or blue edges in the graph is an atomic mutation between the pair of predicates; the red edge is a link to be broken, and the blue edge is a link to be established in a hypothetical automatic fixing application. {In the example above, we could see that out of 41 edges shown in Figure~\ref{fig:compare}, there are 10 common edges, 20 edges are added, and 11 edges are removed from the faulty model, which outlines the atomic mutations. }

\section{Future Work}

In this section, we present several future improvements to the CSBASG representation and planned applications.

\subsection{Laplacian of CSBASG and Mutant as a Control Operation}
At the beginning of this research, we attempted to use a Laplacian matrix to represent the matrix. However, since some self-edges exist in the scope of the Alloy ASG, we could not as there is not currently a  
Laplacian construction for a directed graph containing self-edges. Nonetheless, technically, the CSBASG still holds its structural balance, and the ability to use mutants as control operations in a discrete-time system still holds. There are existing works \cite{acikmese2015spectrum, PREETHAP2023e17001} that could potentially lead to constructions that could apply to various kinds of graphs with self-loops, which are potentially fitting constructions to apply towards a repair process that uses a pair of code segments as a control system in order to explore additional methods of a modification on a software system containing multiple functional code segments under consideration. 

\subsection{Mitigation of Exponential Growth of Matrix Entry}
Lemma 2 mentions a strict exponential bound $O(2^{pT}) $to represent code segments within a repeat-free graph representation. Unfortunately, both the value of repeating nodes in the AST $T$ roughly increases linearly with the increase in code segment size, and the maximum number of nodes under a category of nodes $p$ could also end up a high constant that can vary between different code segments, all of which can prevent a universal, complete representation. Furthermore, even in our representation schema for an Alloy predicate body, as mentioned in Section 4, a rough number of $p=17$ was assigned for any technically infinitely extensible structures such as a ListExpr. While some intuitive solutions exist, like a linked or nested approach to break an infinitely extensible node into multiple nodes linked with each other, they are also costly for time and space. Therefore, the exponential growth of the adjacency matrix remains a challenge for future improvements. 

\subsection{Automated Repair Using Machine Learning}
Our research was initially motivated by a desire to create an input format for machine learning models that would preserve the logical structure of the underlying language, as current machine learning repair techniques treat code as natural language, which limits their likelihood of generating valid Alloy formulas. 
Since we already have a mechanism that could output the common edges, edges to be removed, and edges to be established, we could begin with a predictive model for the probability of the existence of each edge, as suggested by some state-of-the-art graph neural networks, by adapting those methods on the directed multigraph with a polynomial encoding. 

\subsection{Learning of Encoding and Alloy Code Generation}
In this paper, we only attempted to ensure zero-information-loss encoding for Alloy code segments, but we still relied on a map of nodes since we have a static, same-distance pre-defined encoding. Nevertheless, in theory, we can stop this reliance by ensuring no different pair of edges have the same angular signature difference. 

Consider a vector $\bm{\zeta}=[1\angle\theta_1, 1\angle\theta_2, ..., 1\angle\theta_n]^T$ where $n$ is the number of total nodes, such that, by \textit{Lemma 1}, the eigenvector of the CSBASG. We could train this vector as an embedding of the nodes in the CSBASG, giving their angular signatures. In a future scenario, such a vector could be trained with some objective function in applications such as Alloy code generation. A common approach could be an dual-annealing optimization problem like 
$$\min \sum_{(v_i, v_j, k)\in\mathcal E}|\theta_i-\theta_j|$$
$$\max \sum_{\forall k: (v_i, v_j, k)\notin \mathcal E}|\theta_i-\theta_j|$$
given $\theta_i\neq \theta_j$ for any $i\neq j$ and $(v_i, v_j, k)$ is the $k$-th edge in the multigraph form between $v_i$ and $v_j$, minimizing the signature differences between any pairs of nodes while maximizing the signature differences for the nodes that are not directly connected. For a sub-ASG containing solely expressions or formulae with such a vector $\bm{\zeta}$, we could estimate that it could be a high probability for an incomplete ASG to have edges connecting to the most possible nodes that have a lower angular signature distance with the parent node, helping us to fill out the blanks in an incomplete code segment or generate examples within a predefined declarative environment. 

\subsection{Possible Application on Other Languages}
By construction, every programming language is built on a Context-Free Grammar, which we can utilize to create a complete construction and encoding for a CSBASG that can be used in place of an AST. However, our current completeness representation is built on the assertion that each node has limited extensibility. For real-world applications, there could be plenty of structures that either allow infinite extensions or, in a practical sense, have a value of the maximum links down from a category of node sufficiently high enough to hinder the CSBASG's scalability severely. One way to mitigate this is to focus on building CSBASGs of small-scale, simple, and reusable code segments, such as a function body, and then use this representation to help determine its proximity to any known faulty code segments.

\section{Related Work}

In this section, we give an overview of the related work relevant to our planned future applications of a CSBASG for Alloy.

\textbf{Automatic Repair of Alloy Models.}
Automatically repairing faulty Alloy models is a growing research field~\cite{amra, icebar, bounded, atrrepair, tsra6}. ARepair, ICEBAR, and BeAFix are generating valid repair techniques that involve bounded exhaustive searches~\cite{amra,icebar,bounded}. TAR is a mutation-oriented repair technique designed specifically for Ally4Fun models~\cite{tsra6}. ATR tries to find patches based on a preset number of templates~\cite{atrrepair}. While these techniques try to strike different balances in establishing a domain of patches to search through, the current state-of-the-art ATR can only correct 66\% of the Alloy4Fun benchmark models. In addition to opening up the door to explore contextually aware machine learning repairs, our CSBASG could be incorporated into the backend of these existing techniques to try and improve their scalability, many of which define their search space using ASTs.


\textbf{Code Generation for Alloy Models.}
HiGenA is a hint generator for Alloy4Fun exercises that uses historical edits to suggest a series changes a user can make to get to a correct solution~\cite{de2023data}.  
HiGenA's user study highlights that users only find the hints helpful when their solution is already close to the correct answer, indicating room for improvements for hint generation. 
Overall, while restricted to Alloy4Fun exercise, HiGenA does establish a baseline for us to compare our code generation techniques against.
Similar to hint generation, there is an existing body of work that completes partial Alloy models. ASketch takes as input a partial Alloy model with holes and a test suite that outlines the expected behavior~\cite{asketch}. ASketch then generates substitutions for the holes~\cite{rexgen} and tries to find a completed model that passes all tests. However, despite several advances,  ASketch still times out on the largest benchmark model~\cite{SullivanPhDDissertation,asketchgen}. While ASketch does not currently utilize ASTs, other synthesis strategies could be built utilizing our CSBASG representation and compared to ASketch. 

\section{Conclusion}
This paper introduces a novel graph representation of a programming language as a complex-valued multigraph with a polynomial-based adjacency matrix encoding. First, we present how to create a CSBASG of the declarative specification language Alloy. Then, we evaluate the effectiveness of this representation by investigating the compactness improvements the CSBASG achieves compared to an AST for Alloy models, and the comparability CSBASG enables of Alloy predicates. 
In addition, we point out several future research directions that utilize this representation, including repair, code generation and improvements to the representation itself. 




%

\bibliographystyle{ACM-Reference-Format}
\bibliography{ref}
\end{document}